\documentclass[aps,showpacs,floats,twocolumn]{revtex4}
\usepackage{graphics,graphicx,epsf}

\begin{document}

\title{\bf Quasistatic x-ray speckle metrology of microscopic magnetic
return point memory}

\author{Michael S. Pierce$^{1}$, Rob G. Moore$^{1}$, Larry B.
Sorensen$^{1}$, Stephen D. Kevan$^{2}$,
Jeffrey B. Kortright$^{3}$, Olav Hellwig$^{4}$ and  Eric E. Fullerton$^{4}$}

\address{$^{1}$Department of Physics, University of Washington,
Seattle, Washington 98195, USA}

\address{$^{2}$Department of Physics, University of Oregon, Eugene,
Oregon 97403, USA}

\address{$^{3}$Lawrence Berkeley National Laboratory, Berkeley,
California 94720, USA}

\address{$^{4}$IBM Almaden Research Center,  San Jose, California 95120, USA}

\date{\today}

\begin{abstract}
We have used coherent, resonant, x-ray magnetic speckle patterns to measure
the statistical evolution of the microscopic magnetic domains in
perpendicular magnetic films as a function of the applied magnetic field.  Our
work constitutes the first direct, ensemble-averaged study of microscopic
magnetic return point memory, and demonstrates the profound impact of
interfacial roughness on this phenomenon.  At low fields, the
microscopic magnetic domains forget their past history with an
exponential field dependence.
\end{abstract}

\maketitle


The invention of the laser revolutionized visible optics.
We are on the verge of a similar revolution in x-ray
optics.  Fully-coherent, quasicontinuous beams of x-rays
are now available from the third-generation synchrotron
sources, and extremely intense, fully-coherent, pulsed beams from
fourth-generation synchrotron sources are on the horizon.
One of the most exciting applications is to use these coherent
beams to do the x-ray analog of laser light scattering to study
the spacetime correlations in materials at the nanoscopic and
atomic length scales. Such coherent x-ray studies have achieved
time resolutions from microseconds to nanoseconds at a
length scale of a few nanometers\cite{LC}.  Another promising
application is to use the fourth-generation sources
to do holographic or speckle x-ray imaging
to determine the complete structure of the sample with atomic
resolution using a single femtosecond duration beam pulse. Such
``lenseless speckle reconstruction" has recently  been
demonstrated for x-ray charge scattering\cite{recon}, and it is just
beginning to be explored for x-ray magnetic scattering\cite{magrecon}.

In this letter, we describe a new form of reconstructionless x-ray speckle
metrolology using coherent, resonant, magnetic x-ray scattering.  We show
that the speckle patterns act as a fingerprint of the specific domain
configuration and allow the ensemble of microscopic magnetic domains to be
monitored versus the applied field history.  By comparing speckle
patterns, we obtain a quantitative measure of the domain evolution during the
reversal process and directly probe the microscopic mechanisms of
hysteresis.   Magnetic hysteresis is fundamental to all magnetic storage
technology, and this technology is one cornerstone of the present
information age.  Yet, despite decades of intense study\cite{books}
and significant recent advances\cite{sethna}, we still do not have a
fully satisfactory microscopic understanding of magnetic hysteresis.

In his 1905 dissertation at Gottingen, Madelung
defined macroscopic return point memory (RPM)
as follows: Suppose a magnetic system on
the major hysteresis loop is subject to a change in the
applied field that causes an excursion along a minor
hysteresis loop inside the major loop; if the applied
field is readjusted back to its original value and
the sample returns to its initial magnetization, then
macroscopic RPM is said to exist.  Madelung's macroscopic
characterization immediately raises the question of how
the ferromagnetic domains behave on a microscopic level.
Do the domains remember (i.e., return precisely to) their
initial states, or does just the ensemble average remember?
In agreement with previous qualitative studies\cite{prev},
we find that our samples have perfect
macroscopic RPM, but that they have imperfect
microscopic RPM. In addition, we quantitatively measure the
fraction of the domains that remember versus the applied
field history and show that the roughness has a profound
impact on the microscopic RPM.

In the past, microscopic RPM has primarily been studied using
Barkhausen noise\cite{bark}.  For some systems, there are extremely
clear repetitions of the Barkhausen noise that indirectly indicate
that there is microscopic RPM.  However, other systems show
only limited (or no) repetition.  Barkhausen noise measurements
indirectly probe the ensemble of the magnetic domains via the
measured time structure of the spin-flip avalanches. In contrast,
we use coherent x-ray magnetic scattering  to directly probe
the microscopic structure of the domains via the dynamic
structure factor $S(\bf{Q},\omega)$, which is the
space-time Fourier transform of the two-point correlation function
of the magnetization density $\langle M({\bf{r}},t) M({\bf{0}},0) \rangle$
inside the coherently illuminated region.

Our samples were grown on smooth, low-stress,
150-nm-thick $\rm {SiN_x}$
membranes by magnetron sputtering at different argon gas pressures.
The multilayer structure of our two samples is identical
$$\rm{ Pt~(20 ~nm) \ [Co~(0.4 ~nm)/Pt~(0.7 ~nm)]_{50} \ Pt~(1.5 ~nm) } ,$$
\noindent
but their interfacial roughness increases with increased sputtering
pressure.  The sputtering pressure modifies the energies of the deposited
atoms and alters the growth kinetics.  In general, low-pressure sputtering
results in smooth layers while high-pressure growth leads to cumulative
roughness that evolves into domed columns with well defined grain
boundaries [8].  Our goal is to study the microscopic
RPM versus the roughness.
The two samples discussed here were grown at 3 and 12
mTorr with surface roughness, as determined by atomic force microscopy, of
0.45 and 0.90 nm rms, respectively.  The major hysteresis loops measured
with the applied field perpendicular to the film reflect the difference in
the roughness and are shown in Fig. 1.
The smooth, low-pressure sample (3 mT) exhibited reversal
by nucleation and domain wall motion and the shape of its major
loop reflects the thin film geometry and the perpendicular
anisotropy\cite{kooy}.
The rough, high-pressure sample (12 mT) exhibited
much higher coercive fields as a direct consequence of the
increased interfacial roughness.
Kerr and SQUID measurements showed that both of
our samples exhibited ``perfect" macroscopic RPM.

\begin{figure}
\epsfxsize=8.0cm
\epsffile{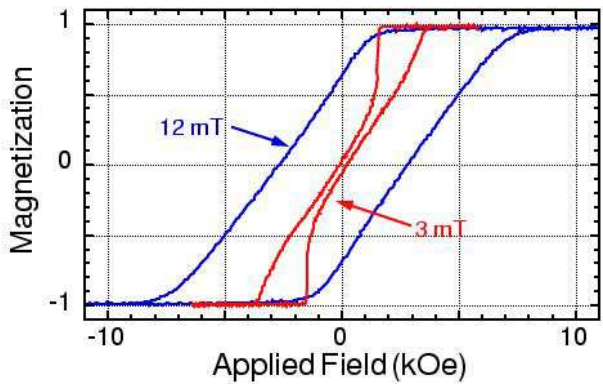}
\caption{(color online) Kerr measurements of the major loops
for the samples grown at different argon sputtering pressures.}
\end{figure}

Our experiments, performed at the Advanced Light Source at Lawrence
Berkeley National Laboratory, use linearly polarized x-rays from
the third harmonic of the beam line 9 undulator\cite{bojeff}.
The photon energy is set to
the cobalt $\rm{L_3}$ resonance at $\sim 778$ eV.  To achieve transverse
coherence,  the raw undulator beam is passed through a 25-micron-diameter
pinhole
before  being scattered in transmission by the sample.  The resonant magnetic
scattering  is collected by a soft x-ray CCD camera.  The magnetic domains are
manipulated  by an electromagnet, which applies fields perpendicular to the
film.
The intensity of the raw undulator beam is $\sim2 \times 10^{14}$ photons/sec,
that of the coherent beam is $\sim2 \times 10^{12}$ photons/sec,
and that of the scattered beam is $\sim2 \times 10^7$ photons/sec.
We collect data for 100 seconds at each magnetic field value, so
our speckle patterns have a total of $\sim2 \times 10^9$ photons
in $\sim 10^6$ ccd pixels.

A typical magnetic speckle pattern for the 3 mT sample at zero
applied field is shown in Fig. 2.  The dominant structure
is a ring of diffuse scattering reminiscent of the scattering from
a classical 2d liquid exhibiting short-range positional correlations.
Note that this diffuse scattering is strongly speckled due to
our use of transversely coherent x-rays. Since changes in the domain
structure will produce changes in the speckle pattern, the degree
of microscopic RPM can be precisely measured by cross correlating
different speckle patterns. By converting the auto- and
cross-correlation functions into a normalized correlation coefficient,
we can obtain a precise measure  of the degree of cross-correlation
between the domains without performing an explicit ``speckle
reconstruction." Note that even if we did perform a speckle reconstruction,
or directly measured the magnetic domains using an
imaging technique, we would still need to reduce the total amount of
information down to a reasonable size to produce a compact, convenient,
ensemble-averaged measure of the microscopic RPM. To do so, we have chosen
the simplest scalar measure of the changes in the speckle pattern,
which is given by a correlation coefficient.  As usual, x-ray
diffraction and high-resolution imaging are complementary probes;
diffraction offers a very convenient way to obtain an ensemble
average; imaging provides the best information about individual
domains and defects.

\begin{figure}
\epsfxsize=8.0cm
\epsffile{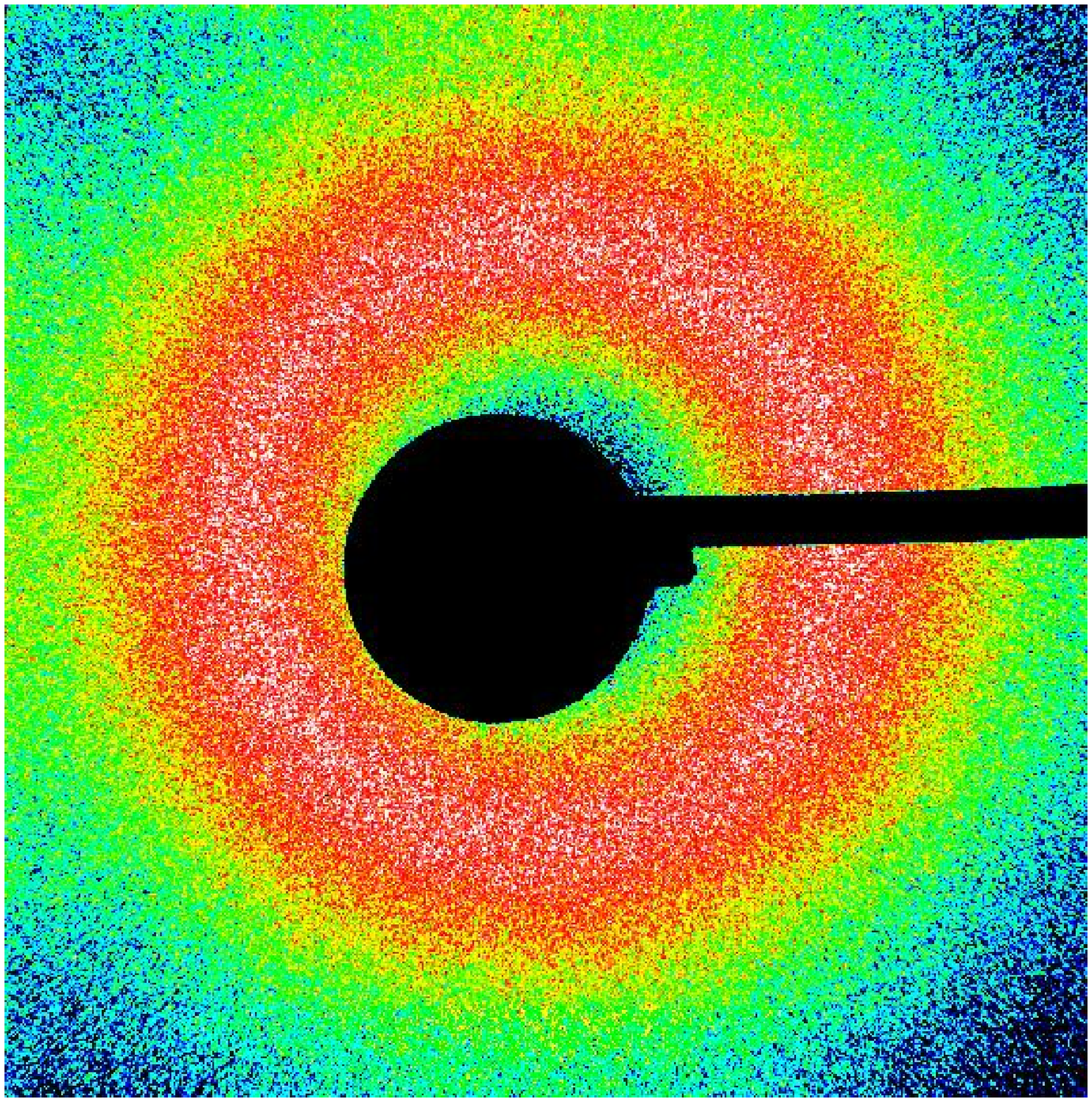}
\epsfxsize=8.0cm
\epsffile{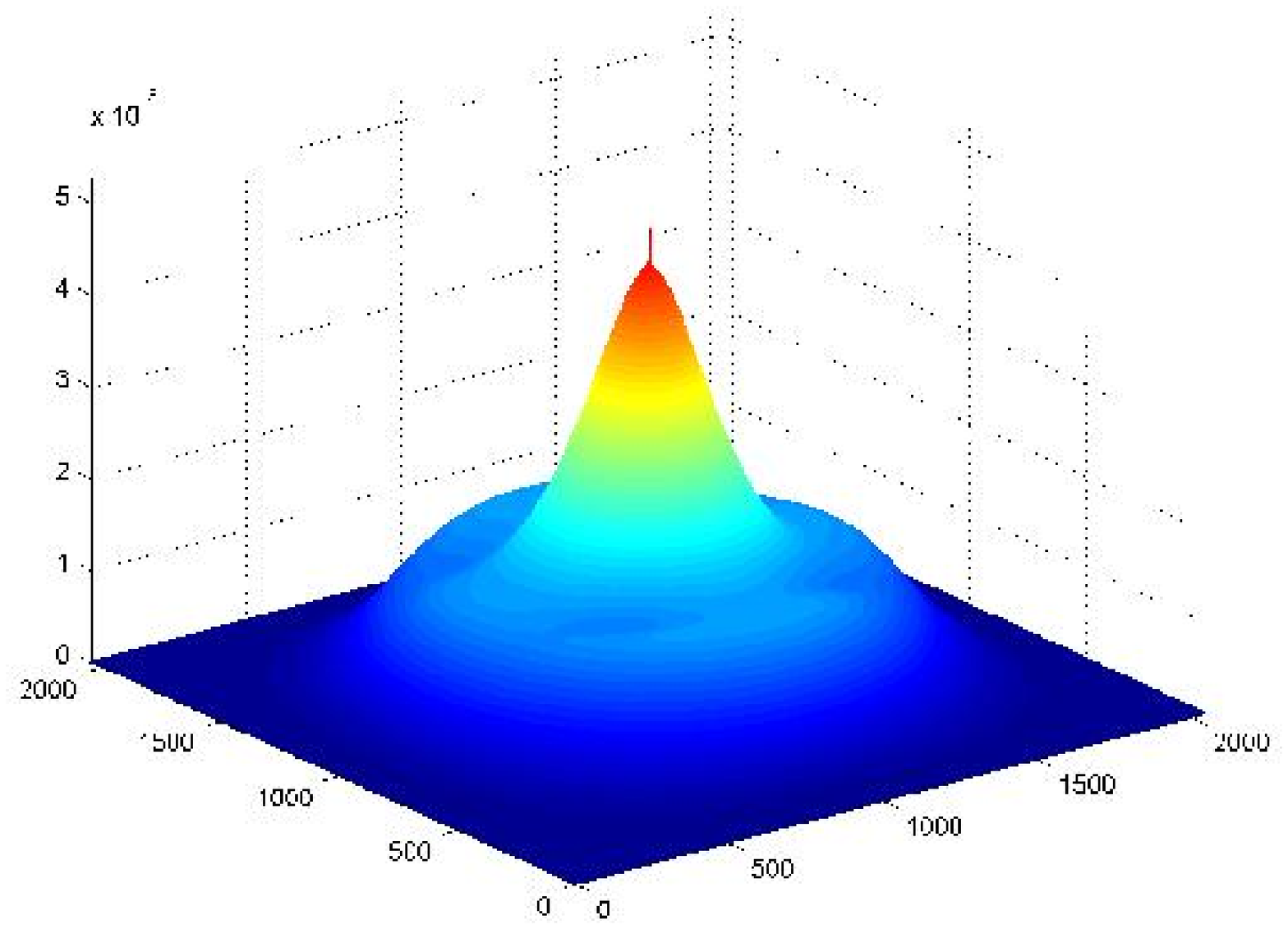}
\caption{(color online) X-ray metrology for the 3 mT sample.
Top: the measured speckle pattern at zero field.
Bottom: the calculated autocorrelation function.
The very sharp spike in the center is due to the
speckles produced by the coherent scattering;
the broad mountain under the coherent spike
is due to the smooth average (incoherent) scattering.}
\end{figure}

To quantify the correlation between two speckle patterns, we
generalized the standard correlation coefficient for two random
variables $a$ and $b$, which is given by

$$\rho(a, b) = {\rm Cov}(a,b) \times \big\{ {\rm Var}(a) {\rm Var}(b)
\big\}^{-1/2} ,$$

\noindent
to the analogous expression in terms of the cross-correlation
function (CCF) and the autocorrelation functions (ACF) of
the speckle pattern intensities $A(q_x,q_y)$ and $B(q_x,q_y)$

\begin{eqnarray} \rho(A, B) & =& \sum \big[{\rm CCF}(A,B)- D \big]
\nonumber \\
& \times& \bigg\{ \sum \big[ \;{\rm ACF}(A)-E \big] \sum \big[
\;{\rm ACF}(B)-F
\big] \bigg\} ^{-1/2} \nonumber
.
\end{eqnarray}

\noindent
In order to collect all of the speckle information,
which is spread over several pixels because our speckles
are 2-3 pixels wide, we first sum over the
pixels near the center of the auto- or cross-correlation
function which contain the speckle information, and we
then subtract the incoherent  background levels (i.e.,
$D(q_x,q_y)$, $E(q_x,q_y)$, and $F(q_x,q_y)$) from each one.
Our generalized correlation  coefficient $\rho$ is unity
for identical speckle patterns, and is zero when the
patterns are completely uncorrelated.  In general, the
value of $\rho$ specifies the degree of correlation between
the two speckle patterns which in turn are proportional to
the Fourier coefficients of the magnetization density for
the two magnetic domain configurations.  A typical example
of one of our experimental  autocorrelation functions is
shown in Fig. 2.

The first question that we addressed was whether our
samples exhibited microscopic RPM through
magnetic saturation.  To do so, we compared two speckle
patterns collected at the same point on the major
loop, but separated either by a minor loop that
returned to saturation or by one or more full
excursions around the major loop. For our smooth
3 mT sample, we found no correlation ($\rho = 0$)
for speckle patterns separated by a return to
saturation.  For our rough 12 mT sample, we found that
there was a repeatable correlation with
$\rho \sim 0.2 - 0.6$ between the speckle
patterns at a given magnetic field separated by
one, or even by many, complete major loops.

Since our smooth 3 mT sample did not have microscopic
RPM through saturation, we measured the decay
of its microscopic RPM as it approached
magnetic saturation to determine how it forgets.
Our results are shown in Fig. 3.
We measured five speckle patterns which were
prepared using the following magnetic field
history protocol:
(1) negative saturation, (2) field $H_1 = H$,
(3) field $H_2 = H$,
(4) field $H_3 = 0$  and (5) field $H_4 = H$.
The negative saturation speckle pattern was used
to determine the background scattering;
$\rho(H_1, H_2)$ was used to test for
any slow time dependence of the speckle patterns
at that field value;  $\rho(H_2, H_3)$
was used to determine the irreversible component of the domain
magnetization---this is analogous to the standard macroscopic
measurement of the remanance magnetization to separate the reversible
and irreversible components of the macroscopic magnetization;
$\rho(H_2, H_4)$
was used to determine the total (reversible plus irreversible)
component of the domain magnetization.
We calculated the normalized reversible and irreversible
fractional changes via
$F(\rm{rev}) = [\rho (H_2, H_4) - \rho (H_2, H_3)]/\rho(H_1, H_2)$
and
$F(\rm{irrev}) = [\rho (H_1, H_2) - \rho (H_2, H_3)]/\rho(H_1, H_2)$.

Figure 3 shows two different kinds of relaxation of the
magnetic domains (1) a field-dependent, quasistatic relaxation
at low fields, and (2) a time-dependent, field-dependent
relaxation at high fields.
Below 2 kOe, $\rho(H_1,H_2)$ remains near unity and both
$\rho(H_2,H_3)$ and $\rho(H_2,H_4)$ show a field-dependent,
quasistatic, exponential decrease. This confirms
the absence of instrumental drift in our measurements and
shows that the magnetic domain structure is static during
the several minutes required to collect the data.
Above 2 kOe, the monotonic decrease in $\rho(H_1,H_2)$
reveals a time-dependent, field-dependent relaxation of
the magnetic domains.
Note that the reversible and irreversible fractional changes
are essentially zero near zero field, and that both increase
for higher field values.

\begin{figure}
\epsfxsize=8.0cm
\epsffile{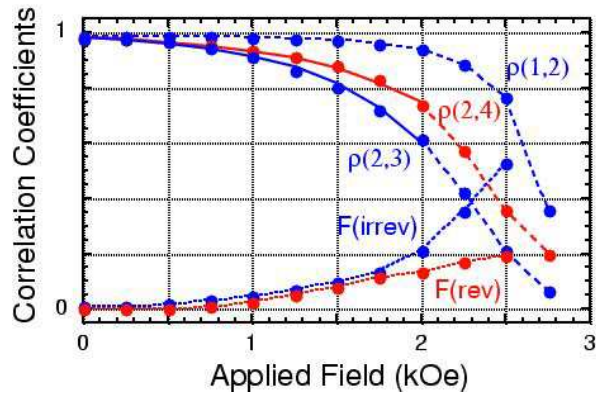}
\caption{(color online) The measured minor loop
microscopic correlation coefficients
for the 3 mT sample.  The solid lines represent fits to
exponential decays.  The dashed lines are guides to the eye.}
\end{figure}

For our rough 12 mT sample, we measured the
correlation coefficient for pairs of points
at the same applied field value on the
major loop which were separated by either one complete
major loop or by eleven complete major loops.
As shown in Fig. 4, this sample exhibits repeatable
major loop microscopic RPM with $\rho \sim 0.2 - 0.6$
depending on the applied field value.
This demonstrates that sufficient interfacial roughness
produces partial major loop microscopic RPM.
We find the highest $\rho$ values near the nucleation
point for reversal and a systematic decrease in $\rho$
toward saturation.  This suggests that the same
nucleation points initiate the reversal, but that the
domain structure becomes increasingly uncorrelated
as the reversal progresses because different reversal
pathways are sampled by the system.

Our results are the first detailed, ensemble-averaged measurements
of the decay of the microscopic RPM in magnetic
materials.  For our smooth sample, saturation completely destroyed
the microscopic RPM.  For our rough sample, saturation only
partially destroyed the microscopic RPM.
The theoretical studies of microscopic RPM
have focussed on the behavior of zero-temperature random field
Ising models (RFIM) which have perfect major and minor loop
microscopic RPM\cite{sethna}.
It will be very interesting to see if the exponential
decays of the microscopic RPM that we find in
our experiments emerge in a natural way from
either the finite-temperature RFIMs\cite{sethna} or the
theories for perpendicular magnetic systems\cite{kooy}.

\begin{figure}
\epsfxsize=8.0cm
\epsffile{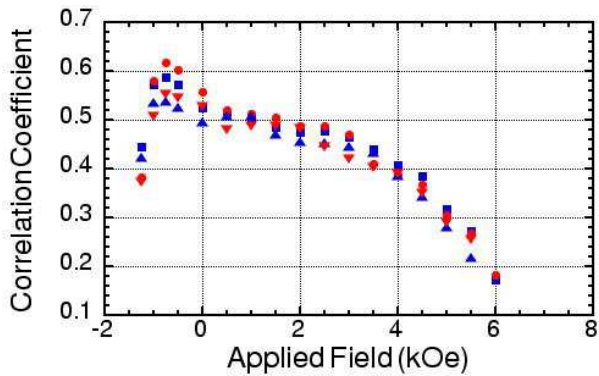}
\caption{(color online) The measured major loop microscopic
correlation coefficients for the 12 mT sample.
The circles and squares (the triangles) represent the
correlations between the first and the second (eleventh) loop.}
\end{figure}

In addition to ferromagnets, there is a wide variety
of materials that exhibit hysteresis and memory effects for which
we do not yet have a satisfactory microscopic understanding, for example
ferroelectrics\cite{ferroe}, spin glasses\cite{spinglass}, and shape memory
materials\cite{shapemem}.  The coherent x-ray scattering cross-correlation
method that we have developed can also be used to study the
microscopic memory in these systems. In addition to the quasistatic
behavior that we have addressed here,
coherent x-ray magnetic scattering
can also be used to study the incoherent,
thermally-driven dynamics of the microscopic domains\cite{LC}.
\begin{acknowledgments}
We gratefully acknowledge the support of our work by the U.S. DOE
via DE-FG06-86ER45275, DE-AC03-76SF00098, and DE-FG03-99ER45776.
O.H. was partially supported by the Deutsche Forschungsgemeinschaft
via HE 3286/1-1.
\end{acknowledgments}


\begin{references}

\bibitem{LC}
A.C. Price {\it et al.}, Phys. Rev. Lett. {\bf 82}, 755  (1999);
I. Sikharulidze {\it et al.}, Phys. Rev. Lett. {\bf 88}, 115503 (2002).

\bibitem{recon}
See for example:
J. Miao and D. Sayre,  Acta Crystallog. A {\bf 56}, 596 (2000);
I.A. Vartanyants and I.K. Robinson,  J. Phys. C {\bf 13},10593 (2001);
U. Weierstall {\it et al.}, Ultramicros. {\bf 90},171 (2002);
and references therein.

\bibitem{magrecon}
T.O. Mentes, C. Sanchez-Hanke and  C.C. Kao, J. Sync, Rad. {\bf 9}, 90 (2002);
F. Yakhou {\it et al.}, J. Magn. Magn.  Mat. {\bf 233}, 119 (2001);
A. Rahmim, {\it Analysis of Coherent Resonant X-Ray Scattering and
Reconstruction of Magnetic Domains}, M.Sc. Thesis, University of
British Columbia, 2001.

\bibitem{books}
See for example:
G. Bertotti, {\it Hysteresis in Magnetism} (Academic, San Diego, 1998);
M. Brokate and J. Sprekels, {\it Hysteresis and Phase Transitions}
(Springer, New York 1996); E. Della Torre, {\it Magnetic Hysteresis}
(IEEE Press,New York, 1999).

\bibitem{sethna}
See for example:
J.P. Sethna, K.A. Dahmen and C.R. Myers, Nature (London)
{\bf 410}, 242 (2001); J.H. Carpenter {\it et al.},
J. Appl. Phys. {\bf 89}, 6799 (2001); K.A. Dahmen {\it et al.},
J. Magn. Magn. Mat. {\bf 226-230}, 1287 (2001); K.A.
Dahmen, J.P. Sethna and O. Perkovic, IEEE Trans.
Magn. {\bf 36}, 3150 (2000); and references therein.

\bibitem{prev}
See for example: H.S. Cho {\it et al.}, IEEE Trans. Magn. {\bf 34}, 1150
(1998);
A. Hulbert and R. Schaefer, Sec. 5.6.1 in {\it Magnetic Domains}
(Springer, Berlin, 1998);
and P. Fischer {\it et al.}, J. Phys. D {\bf 31}, 649 (1998).

\bibitem{bark}
S. Zapperi  and  G. Durin, Comp.
Mat. Sci. {\bf 20}, 436 (2001);
J.R. Petta, M.B. Weissman and  G. Durin,
Phys. Rev. E {\bf 56}, 2776 (1997);
D. Spasojevic {\it et al.}, Phys.
Rev. E {\bf 54}, 2531 (1996).

\bibitem{eric}
E.E. Fullerton {\it et al.}, Phys. Rev. B {\bf 48}, 17432 (1993).

\bibitem{kooy}
C. Kooy and C. Enz, Philips Res. Reports {\bf 15}, 7 (1960);
Y. Yafet and E.M. Gyorgi, Phys. Rev. B {\bf 38}, 9145 (1988);
B. Kaplan and G.A. Gehring, J. Magn. Magn. Mat. {\bf 128}, 111 (1993);
A. Kashuba and V.L. Pokrovsky, Phys. Rev. Lett. {\bf 70}, 3155 (1993)
and Phys. Rev. B {\bf 48}, 10335 (1993)

\bibitem{bojeff}
J.B. Kortright {\it et al.}, Phys. Rev. B {\bf 64}, 092401 (2001);
B. Hu {\it et al.}, Synch. Rad. News {\bf 14},11 (2001).


\bibitem{ferroe}
E.V. Colla, L.K. Chao  and M.B. Weissman,   Phys. Rev. Lett.
{\bf 88}, 17061 (2002) and Phys. Rev. B {\bf 63}, 134107 (2001).

\bibitem{spinglass}
M.B. Weissman,  Physica D {\bf 107}, 421
(1997) and Rev. Mod. Phys. {\bf 65}, 829 (1993).

\bibitem{shapemem}
L. Carrillo and  J. Ortin,   Phys. Rev. B {\bf 56}, 11508 (1997).


\end{references}
\end{document}